\UseRawInputEncoding 

\documentclass[twocolumn,showpacs,superscriptaddress,aps,pra]{revtex4-1}
\usepackage{epsfig}
\usepackage{subfigure}
\usepackage{amssymb}
\usepackage{graphicx}
\usepackage{color}
\usepackage{amsfonts}
\usepackage{amscd}
\usepackage{hyperref}
\usepackage{amsmath}
\usepackage{epstopdf}

\usepackage{amsmath}
\usepackage{graphicx}
\usepackage{txfonts}
\usepackage{textcomp}

\begin{document}

\title{Trade-off relations of quantum resource theory in neutrino oscillations}

\author{Yu-Wen Li}

\author{Li-Juan Li}

\author{Xue-Ke Song}
\email{songxk@ahu.edu.cn}

\author{Dong Wang}
\email{dwang@ahu.edu.cn}

\affiliation{School of Physics and Optoelectronics Engineering, Anhui University, Hefei 230601, China}

\date{\today }

\begin{abstract}

The violation of the classical bounds imposed by Leggett-Garg inequalities has tested the quantumness of neutrino oscillations (NOs) over a long distance during the propagation.
The measure of quantumness in experimentally observed NOs are studied via quantum resource theory (QRT).
Here, we focus on the trade-off relations of QRT in the three-flavor NOs, based on Bell-type violations, first-order coherence and intrinsic concurrence,
and the relative entropy of coherence. For the electron and muon antineutrino oscillations, the analytical trade-off relations
obeyed by the Bell-CHSH inequality of pairwise flavor states in this three flavor neutrino system are obtained; the sum of the maximal violation of
the $\rm CHSH$ tests for three pairwise flavor states is less than or equal to 12. Moreover, there exists an equality relation concerning first-order coherence and
intrinsic concurrence in NOs, showing how much quantum resources flow between first-order coherence and intrinsic concurrence during the neutrino propagation.
In addition, it is found that the tripartite coherence of three-flavor system is equal to or larger than the sum of the coherence of reduced bipartite flavor states.
The trade-off relations of QRT provide a method for studying how the quantum resources convert and distribute in NOs, which might inspire the future applications
in quantum information processing using neutrinos.
\end{abstract}
\maketitle
\section{Introduction}

Neutrino is a fermion with a tiny mass, and interacts only with other particles of matter
by the weak subatomic force and gravity \cite{kuo1989,Garcia2003}. This feature of weak interaction makes it a good candidate to probe environments
that other radiation cannot penetrate. In the neutrino framework, neutrino comes in three flavours, or types, referred to as electron, muon and tau \cite{Vernon}. Maybe it has the forth one, sterile neutrino, but
it is not detected yet in experiment. The flavor states of neutrinos are linear combination of the mass eigenstates~\cite{Camilleri:2008zz,Duan:2010bg}.
 Neutrino oscillation (NO) describes that the three flavors are able to oscillate during flight, i.e.,
flip from one type to another. The measure and analysis of oscillations parameters are great theoretical and experimental interest \cite{Minakata,dgm,abs,PAda,fpa2,Forero}.
 NO guarantees that one can measure
the probability of a particular flavor at the arbitrary time in neutrino propagation, and then can be used to study the classical
or quantum properties of the oscillation process. More broadly, it
has potential applications in probing the unknown object beyond the solar system and future communication.

The Leggett-Garg inequality (LGI), referred to as the "time analogue" of Bell's inequality, tests the correlations of a single system measured at different times. Violation of a LGI implies either the absence of a realistic description of the system or the impossibility of measuring the system without disturbing it. It is shown that
the experimentally observed neutrino oscillations can violate the classical limits imposed by the LGI \cite{qfx,dgd,jaf,dga}. For example, in 2016, Formaggio \emph{et al.} \cite{jaf} presented the classical limits of NOs can be violated by the LGI over a distance of $735~\rm km$, using the data gathered by the MINOS neutrino experiment.
In 2017, Fu \emph{et al.} \cite{qfx} verified the values of LGI $K_3$ and $K_4$ are violated with the quantum mechanics prediction at a confidence level of over $6\sigma$
with the updated data of Daya Bay experiment.
These violations of LGI can act as a new affirmation of quantum nonclassical features existing in the neutrino system in the long-distance propagation.

Recently, three-flavor NOs are treated as a three-qubit quantum system. In particular, the phenomena of particle mixing and flavor oscillations in elementary-particle
physics can be described in terms of multi-mode entanglement of single-particle states, based on flavor transition probabilities \cite{mbf08,mbf09}. Also, in some cases,
they can be reduced to effective two-flavor NOs \cite{aka,jha}. Therefore, we can use the tools of quantum resource
theory (QRT) to study the information-theoretic quantification and performance of quantum correlations in the three-flavor NOs \cite{sba,ljlf,bmd,mbf14,dwfx,xxs,ming1}.
For example, in 2015, Banerjee \emph{et al.} \cite{sba} studied the quantum-information theoretic quantities of three-flavor NOs, including the flavor entropy, Mermin inequalities, Svetlichny inequalities, and quantum discord.  In 2021, Blasone \emph{et al.} \cite{bmd} proposed a wave-packet approach to study nonlocal advantage of quantum coherence and Bell nonlocality of quantum correlations in NOs. In 2021, Li \emph{et al.} \cite{ljlf} characterized
quantum resources originating from NO systems by entanglement of formation, negativity, concurrence and uncertainty relations.

A unified and rigorously defined framework of QRT can help us to investigate many related issues, such as the
characterization, the quantification, the manipulation of quantum states under the imposed constraints \cite{idaw,nhajho}, and the trade-off relations among different measures of quantum
correlations and coherence \cite{yyx,qhh,xyh}. Particularly, the trade-off relations in QRT is of importance for establishing the quantifications and classifications of
quantum correlations in multipartite systems. Based on these relations, one can establish a bound
on a physical quantity via other complementary quantities, and can also deeply understand the intrinsic relationship among different quantum resources \cite{yang2013,xgf,fmw}. Since
QRT has many potential applications in quantum information processing, for example, quantum
key distribution \cite{qkd1,qkd2}, quantum teleportation \cite{qt}, and quantum randomness generation \cite{qrg1,qrg2},
the investigations concerning QRT have attracted increasing attention in both theory and experiment \cite{yyx,qhh,bhatti,wwjh,yanghuan}.
For instance, in 2015, Yao \emph{et al.} \cite{yyx} made an exact relationship between quantum coherence
with other measures of quantum correlations, such as quantum entanglement or quantum discord, in multipartite systems.
In 2019, Wang \emph{et al.} \cite{wwjh} experimentally demonstrated demonstrated and
quantified quantum resource conversion between quantum coherence and quantum discord in the deterministic quantum computation with one qubit model,
based on one pure superconducting qubit.

In this paper, we pay attention to explore the trade-off relations, including Clauser-Horne-Shimony-Holt ($\rm CHSH$) nonlocality inequality, first-order coherence and intrinsic concurrence,
the relative entropy of quantum coherence, in the three-flavor NO systems.
It is found that the trade-off relation is valid that the sum of the maximal violation of the $\rm CHSH$ inequality tests for three pairs of reduced bipartite neutrino flavor states
is less than or equal to 12, suggesting that if one of them is very close to the maximal violation of the $\rm CHSH$
inequality, the other two cannot violate it any more.
For first-order coherence and intrinsic concurrence, the relation that the square of first-order coherence plus two-third of the square of intrinsic concurrence
equals to 1 holds for NOs. In addition, we obtain how the quantum coherence is distributed among the
subsystems of the three-flavor neutrino propagating systems.
The result shows that trade-off relation can be used as a credible criterion for the transfer of quantum resources in the three-flavor NOs.
This could open the window for exploring the practical applications of neutrinos in QRT.

The main paper is organized as follows. In Sect. 2, we briefly introduce three-flavor NOs model. In Sect. 3, we give a review about
the trade-off relations of the $\rm CHSH$ inequality, first-order coherence and intrinsic concurrence, and  relative entropy of coherence. In Sect. 4, we
study those trade-off relations of multipartite systems, based on the electron and muon antineutrino oscillations.
Finally, we end up our article with conclusions in Sect. 5.

\section{Three-flavor NOs}

The three-flavor neutrino model includes three distinct neutrino flavors, $\left| {{\nu _e}} \right\rangle$, $\left| {{\nu _\mu}} \right\rangle $, and $\left| {{\nu _\tau }} \right\rangle $.  And the neutrinos of definite flavor can be written as a quantum superposition of the mass eigenstates, $\left| {{\nu _1}} \right\rangle $, $\left| {{\nu _2}} \right\rangle$, and $\left| {{\nu _3 }} \right\rangle $. That is
\begin{align}
\left| {{v_\alpha }} \right\rangle  = \sum\limits_j {U_{\alpha j}^ * } \left| {{v_j}} \right\rangle,
\label{Eq.1}
\end{align}
where $\alpha  = e, \mu, \tau, $ $j = 1, 2, 3$, and $U_{\alpha j}^ *$ is the complex conjugate of the $\alpha j-$th elements of the $3\times3$ unitary matrix $U$,
which is also referred to as the Pontecorvo-Maki-Nakagara-Sakata matrix.
The matrix $U$ and its decomposition can be expressed as \cite{zmm,ams2017}
\begin{eqnarray}    
\left( {\begin{array}{*{20}{c}}
{{U_{e1}}}&{{U_{e2}}}&{{U_{e3}}}\\
{{U_{\mu 1}}}&{{U_{\mu 2}}}&{{U_{\mu 3}}}\\
{{U_{\tau 1}}}&{{U_{\tau 2}}}&{{U_{\tau 3}}}
\end{array}} \right)&=& \left( {\begin{array}{*{20}{c}}
1&0&0\\
0&{{c_{23}}}&{{s_{23}}}\\
0&{ - {s_{23}}}&{{c_{23}}}
\end{array}} \right)\left( {\begin{array}{*{20}{c}}
{{c_{13}}}&0&{{s_{13}}{e^{ - i{\delta _{cp}}}}}\\
0&1&0\\
{ - {s_{13}}{e^{i{\delta _{cp}}}}}&0&{{c_{13}}}
\end{array}} \!\!\!\right)\nonumber\\
            &&\times\left( {\begin{array}{*{20}{c}}
{{c_{12}}}&{{s_{13}}}&0\\
{ - {s_{12}}}&{{c_{12}}}&0\\
0&0&1
\end{array}} \right)\left( {\begin{array}{*{20}{c}}
{{e^{i{\alpha _1}/2}}}&0&0\\
0&{{e^{i{\alpha _2}/2}}}&0\\
0&0&1
\end{array}}\! \right),\nonumber\\
\label{Eq.2}
\end{eqnarray}
where ${c_{ij}}=\cos{\theta _{ij}}$ and ${s_{ij}} = \sin {\theta _{ij}}$ ($i, j=1, 2, 3$) with $\theta _{ij}$
being the mixing angle, and $\alpha _1$, $\alpha _2$, and $\delta _{cp}$ are CP-violating phases.
The state ${\rm{ }}\left| {{v_j}} \right\rangle $ is the mass eigenstate of the free Dirac Hamiltonian,
\begin{align}
{H_j} = \overrightarrow \xi \cdot {\widehat {\overrightarrow {p_j} }}c + \eta {m_j}{c^2},
\label{Eq.3}
\end{align}
where $c$ is the the speed of the light in free space, ${m_j}$ is the mass, ${\widehat {\overrightarrow {p_j} }}$ is momentum operator, $E_j=\sqrt{|\overrightarrow {k_j}|^2c^2+m^2_jc^4}$.
When the definite flavor of a neutrino propagates along the $x$ axis, then we can write the flavor of propagation by the plane wave solution of the form
\begin{align}
\langle \overrightarrow x|\left. {{v_j }(t)} \right\rangle=\exp\left[-\frac {i}{\hbar}(E_jt-\overrightarrow {k_j}\cdot\overrightarrow x)\right]\langle \overrightarrow x|\left. {{v_j(0) }} \right\rangle,
\label{Eq.4}
\end{align}
where $t$ is the time of propagation, $\overrightarrow {k_j}$ is the momenta, and $\overrightarrow x$ is the position of the particle in the mass eigenstate from the source point.
The neutrino has been shown to have a tiny mass, and hardly interacts with other particles of matter. Thus, in the case of ultra-relativistic neutrinos ($|\overrightarrow {k_j}|\gg m_j c$), its energy can be approximated as
\begin{eqnarray}
E_j=\sqrt{|\overrightarrow {k_j}|^2c^2+m^2_jc^4}\simeq|\overrightarrow {k_j}|c+\frac{m^2_jc^3}{2k_j}\simeq E+\frac{m^2_jc^4}{2E},
\label{Eq.5}
\end{eqnarray}
where $E\approx|\overrightarrow {k_j}|c$, and the same for all $j$. When a neutrino beam $v_\alpha$ is created in a
charged current interaction, the evolved state with respect to time $t$ takes the form of
\begin{eqnarray}
\langle \overrightarrow x|\left. {{v_\alpha }(t)} \right\rangle&=&\sum\limits_jU^\ast_{\alpha j}\exp\left[-\frac {i}{\hbar}(E_jt-\overrightarrow {k_j}\overrightarrow x)\right]\langle \overrightarrow x|\left. {{v_j }} \right\rangle \nonumber\\
&=&\sum\limits_\beta\sum\limits_jU^\ast_{\alpha j}\exp\left[-\frac {i}{\hbar}(E_jt-\overrightarrow {k_j}\overrightarrow x)\right]U_{\beta j}\langle \overrightarrow x|\left. {{v_\beta }} \right\rangle.\nonumber\\
\label{Eq.6}
\end{eqnarray}
In the one-dimensional case, choosing $\overrightarrow {k_j}=(k, 0, 0)$, and the same for all the other $j$, we get
\begin{eqnarray}
\langle k|\left. {{v_\alpha }(t)} \right\rangle&=&\sum\limits_\beta\sum\limits_jU^\ast_{\alpha j}\exp\left[-\frac {i}{\hbar}E_jt\right]U_{\beta j}\langle k|\left. {{v_\beta }} \right\rangle,
\label{Eq.7}
\end{eqnarray}
where $\beta  = e, \mu , \tau $. Based on the assumption that neutrino are ultra-relativistic particles, using
Eq.(\ref{Eq.5}), the amplitude of finding the flavor state $\left| {{v _\beta}} \right\rangle$ in the original $\left| {{v _\alpha}} \right\rangle$ beam at time $t$ is expressed as
\begin{eqnarray}
{{a_{\alpha \beta }}(t)}=\langle v_\beta|\left. {k} \right\rangle \langle k|\left. {{v_\alpha }(t)} \right\rangle\!\!=\!\!\sum\limits_jU^\ast_{\alpha j}\exp\left[-im^2_jc^3
\frac{L}{2E\hbar} \right]U_{\beta j},
\label{Eq.8}
\end{eqnarray}
where $L\simeq ct$ is the traveled distance of the neutrino particle traveled between source and detector.
Finally, the transition probability ${v_\alpha }(t=0) \to {v_\beta }(t)$ is obtained as
\begin{eqnarray}
  P({v_\alpha } \to {v_\beta }) &=&{\delta _{\alpha \beta }} - 4\sum\limits_{j>r} Re(U_{\alpha j}^*{U_{\beta j}}U_{\alpha r}{U_{\beta r}^*}) \nonumber\\
  && \times {\sin ^2}(\vartriangle m_{jr}^2\frac{{L{c^3}}}{{4E\hbar }}) \nonumber\\
  &&+ 2\sum\limits_{j>r} I m(U_{\alpha j}^*{U_{\beta j}}U_{\alpha r}{U_{\beta r}^*}) \nonumber\\
   && \times {\sin}(\vartriangle m_{jr}^2\frac{{L{c^3}}}{{2E\hbar }}),
\label{transition}
\end{eqnarray}
where $\Delta m_{jr}^2 = m_j^2 - m_r^2$. To investigate the trade-off relations of quantum resource theory in NOs, it is convenient to write the oscillatory quantity of Eq. (\ref{transition}),
${\sin}(\vartriangle m_{jr}^2\frac{{L{c^3}}}{{4E\hbar }})$, in a simple form
\begin{eqnarray}
\sin^2\left(\Delta m^2_{jr}\frac{Lc^3}{4E\hbar}\right)=\sin^2\left(1.27\Delta m^2_{jr}[eV^2]\frac{L[km]}{E[GeV]}\right).
\end{eqnarray}
\begin{figure}
\begin{minipage}{0.45\textwidth}
\centering
\subfigure{\includegraphics[width=8cm]{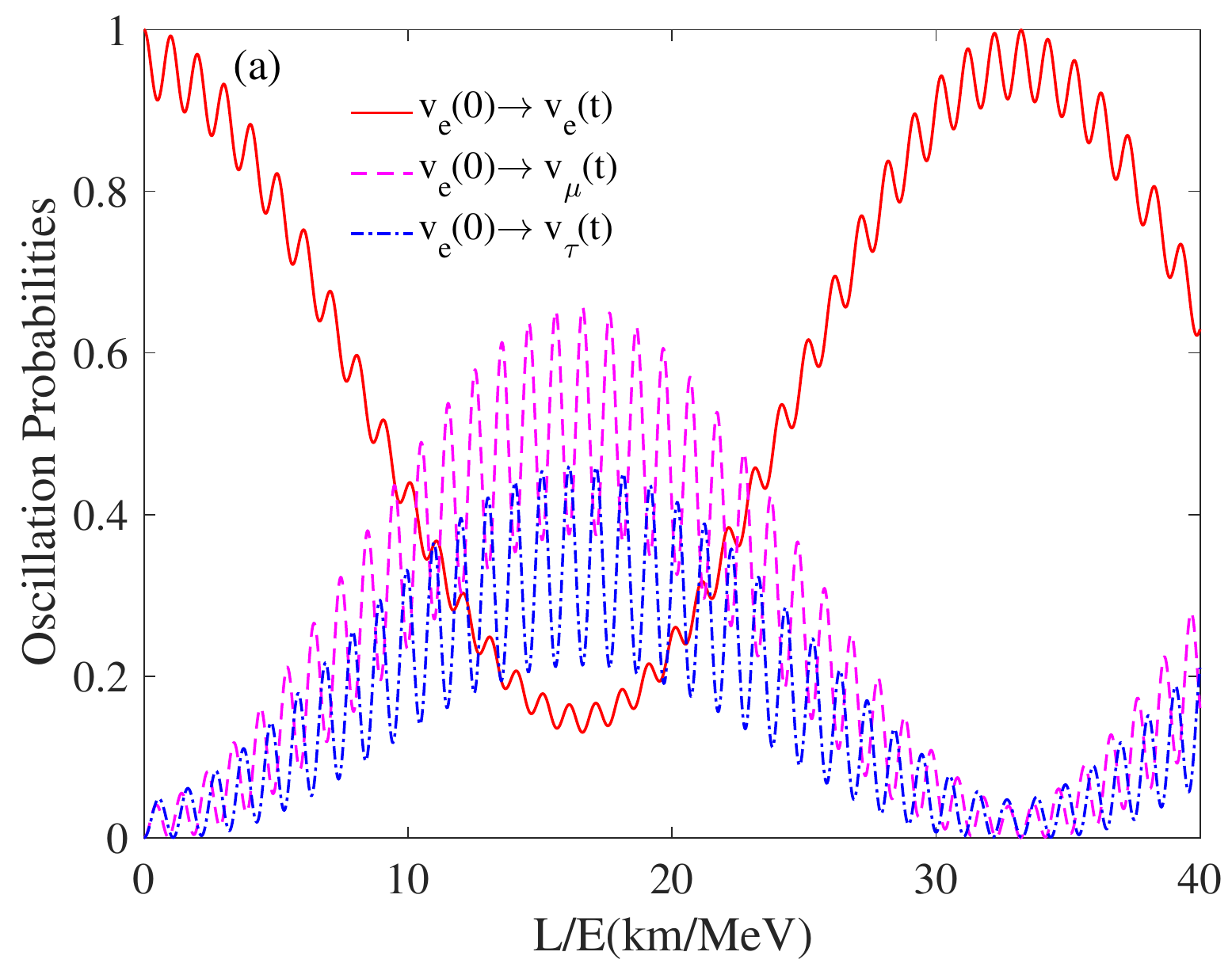}}
\subfigure{\includegraphics[width=8cm]{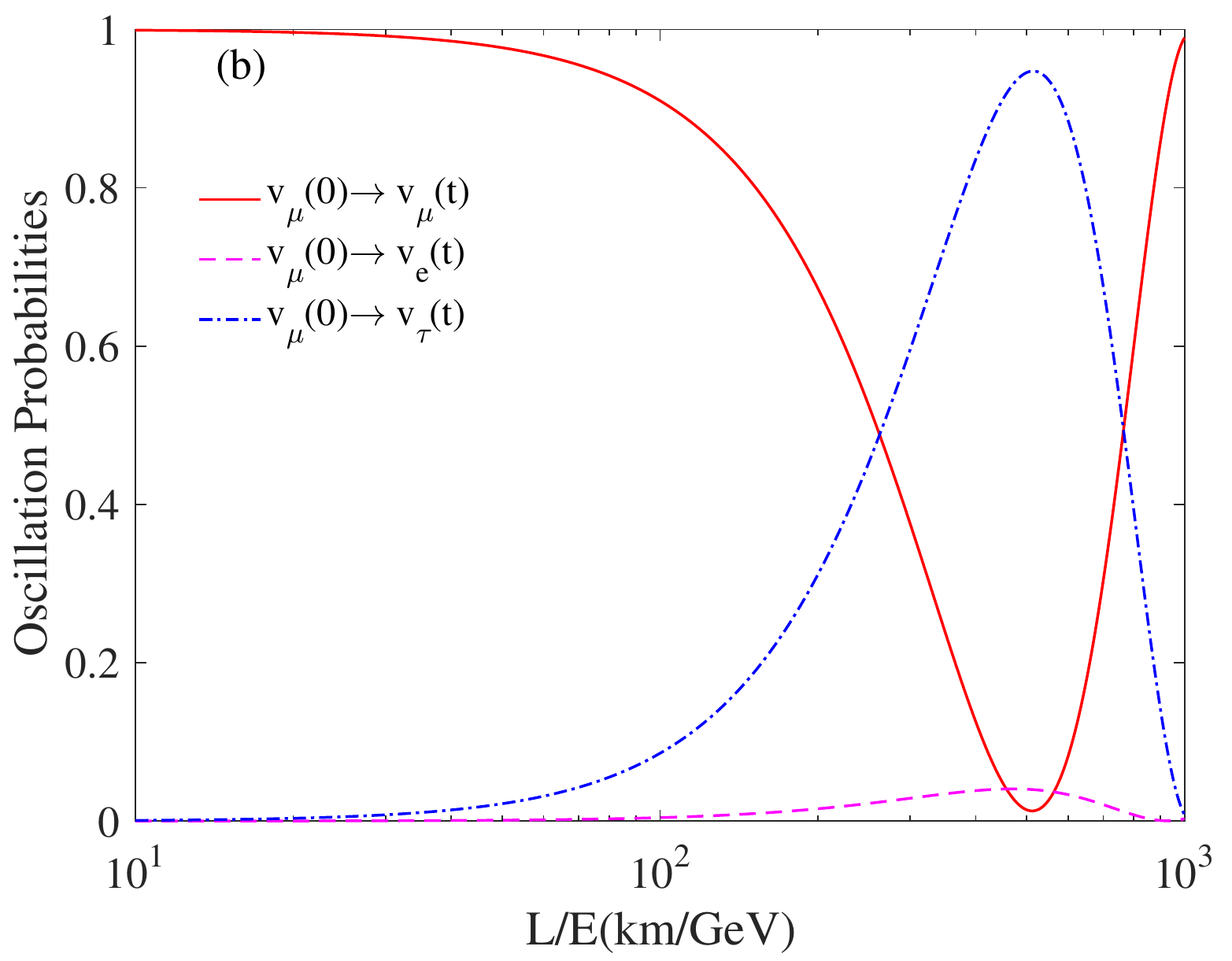}}
\end{minipage}\hfill
\caption{Neutrino oscillation(NO) probability as a function of the ratio between the traveled distance $L$ and the energy $E$. The figure (a) plots the oscillation probability ${v_e}(0) \to {v_e}(t)$ (red, solid line), ${v_e}(0) \to {v_\mu}(t)$ (purple, dashed line), ${v_e}(0) \to {v_\tau}(t)$ (blue, dashed-dotted line) when the initial neutrino flavor is electron flavor. The figure (b) plots the oscillation probability ${v_\mu}(0) \to {v_\mu}(t)$ (red, solid line), ${v_\mu}(0) \to {v_e}(t)$ (purple, dashed line), ${v_\mu}(0) \to {v_\tau}(t)$ (blue, dashed-dotted line) when the initial neutrino flavor is muon flavor. }
\label{f1}
\end{figure}

Also, for normal ordering of the neutrino mass spectrum ($m_1 < m_2 < m_3$), the best
fit values of the three-flavor oscillation parameters are given by
\begin{eqnarray}
\Delta m_{21}^2 &=& 7.50 \times {10^{ - 5}}e{V^2},\\
\Delta m_{31}^2& = &2.457 \times {10^{ - 3}}e{V^2},\\
\Delta m_{32}^2 &=& 2.382 \times {10^{ - 3}}e{V^2},\\
{\theta _{12}} &=& {33.48^ \circ },  {\theta _{23}} = {42.3^ \circ },  {\theta _{13}} = {8.50^ \circ }.
\label{Eq.7}
\end{eqnarray}
As $\delta _{cp}$ has not been  observed by experiments, we can choose $\delta_{cp}=0$ for simplicity.
When the initial neutrino is in electron neutrino, in Fig. \ref{f1} (a), we show the transition probability, ${P_{{\nu _e } \to {\nu _\beta }}}= {\left| {{a_{e \beta }}(t)} \right|^2}$, for the initial flavor $\left| {{\nu _e}} \right\rangle$ to be in $\left| {{\nu _e}} \right\rangle$, $\left| {{\nu _\mu }} \right\rangle$, $\left| {{\nu _\tau }} \right\rangle $ after time $t$, with respect to $L/E$. The survival probability of $\left| {{\nu _e}} \right\rangle$ is always higher than $0.1$, and the probabilities of detecting the other flavors
are smaller than $0.7$ as the variations of $L/E$ from $0$ to $40$ with dimension $\rm km/MeV$.
Fig. \ref{f1} (b) plot the transition probability ${P_{{\nu _\mu } \to {\nu _\beta }}} = {\left| {{a_{\mu \beta }}(t)} \right|^2}$ as a function of $L/E$ when an muon
neutrino is produced at the initial time $t=0$. The survival probability of $\left| {{\nu _\mu}} \right\rangle$ changes from $0$ to $1$, but the probability of detecting the electron
flavor always takes a small value below 0.04 in a range $[10^1, 10^3]$ of $L/E$ with dimension $\rm km/GeV$.

Following the idea of Ref. \cite{mbf09}, the occupation number in the neutrinos can be written as a three-qubit system, which gives
\begin{align}
\begin{array}{l}
\left| {{\nu _e }} \right\rangle  \equiv {\left| 1 \right\rangle _e} \otimes {\left| 0 \right\rangle _\mu } \otimes {\left| 0 \right\rangle _\tau } \equiv \left| {100} \right\rangle,\\
\left| {{\nu _\mu }} \right\rangle  \!\equiv {\left| 0 \right\rangle _e} \otimes {\left| 1 \right\rangle _\mu } \otimes {\left| 0 \right\rangle _\tau } \equiv \left| {010} \right\rangle,\\
\left| {\nu {}_\tau } \right\rangle  \equiv {\left| 0 \right\rangle _e} \otimes {\left| 0 \right\rangle _\mu } \otimes {\left| 1 \right\rangle _\tau } \equiv \left| {001} \right\rangle.
\end{array}
\label{Eq.8}
\end{align}
In this case, the flavor oscillations of neutrino can be seen as the time evolution of a tripartite quantum state. Then, we have
\begin{align}
{\left| {\psi (t)} \right\rangle _\alpha } = {a_{\alpha e}}(t)\left| {100} \right\rangle  + {a_{\alpha \mu }}(t)\left| {010} \right\rangle  + {a_{\alpha \tau }}(t)\left| {001} \right\rangle.
\label{Eq.9}
\end{align}
Therefore, we can discuss the trade-off relation of quantum resource theory, including CHSH inequality, the first-order coherence and intrinsic concurrence, and the relative entropy of
coherence, in this three-qubit quantum system based on NOs.

\section{Trade-off relation of quantum resource theory}

A two-qubit quantum state $\rho$, in terms of the Hilbert-Schmidt representation, can be expressed as
\begin{align}
{\rho}\! =\!  \frac{1}{4}[{I_2} \otimes {I_2} + \mathop a\limits^ \to   \cdot \mathop \sigma \limits^ \to   \otimes {I_2} + {I_2} \otimes \mathop b\limits^ \to   \cdot \mathop \sigma \limits^ \to   + \sum\limits_{i,j} {{m_{ij}}{\sigma _i} \otimes {\sigma _j}}],
\label{Eq.10}
\end{align}
where $I_2$ is the $2\times2$ identity matrix, $\mathop a\limits^ \to$ and $\mathop b\limits^ \to$ are the local Bloch vectors, ${m_{ij}} = \rm Tr(\rho{\sigma _i} \otimes {\sigma _j})$ is the elements of the correlation matrix $M$.

The quantum nonlocality of two-qubit states can be revealed by the violation of Bell-type inequalities, and the best known Bell inequality is the $\rm {CHSH}$ inequality.
The Bell operator corresponding to $\rm {CHSH}$ inequality reads
\begin{align}
\mathcal{B}= {A_1} \otimes {B_1} + {A_1} \otimes {B_2} + {A_2} \otimes {B_1} - {A_2} \otimes {B_2},
\label{Eq.10}
\end{align}
where ${A_i} = \mathop {{a_i}}\limits^ \to  \mathop { \cdot {\sigma _A}}\limits^ \to = a_i^x\sigma _A^1 +a_i^y\sigma _A^2 + a_i^z\sigma _A^3$, $B_j = \mathop {b_j}\limits^ \to\cdot \mathop {\sigma _B}\limits^ \to = b_j^x\sigma _B^1+b_j^y\sigma _B^2 + b_j^z\sigma _B^3$, $\mathop {{a_i}}\limits^\to=(a_i^x, a_i^y, a_i^z)$  and  $\mathop {{b_j}}\limits^ \to = (b_j^x, b_j^y, b_j^z)$ are real unit vectors, and $\sigma _{A/B}^{1,2,3}$ are Pauli matrices. The $\rm {CHSH}$ inequality is then written as
\begin{align}
{\mathbb{B}} =|\langle\mathcal{B}\rangle_\rho|=|\rm Tr(\rho\mathcal{B})|\leq2.
\label{Eq.10}
\end{align}

Let ${\left\langle {\rm CHSH} \right\rangle _\rho}$ denote the maximal mean value ${\left\langle \mathcal{B} \right\rangle _\rho }$ under all possible measurement settings.
Therefore, the ${\left\langle {\rm CHSH} \right\rangle _\rho }$  of a given two-qubit quantum state $\rho $ is given by \cite{mhp}
\begin{align}
{\left\langle {\rm {CHSH}} \right\rangle _\rho } = \mathop {\max }\limits_{{A_i},{B_j}} {\rm{ Tr(}}\rho \mathcal{B}{\rm{) = }}\sqrt {{\tau _1} + {\tau _2}},
\label{Eq.11}
\end{align}
where ${{\tau _1}}$ and ${{\tau _2}}$ are the two largest eigenvalues of the matrix ${M^\dag }M$, and ${M^\dag }$ is the conjugate and transpose of matrix $M$, ${M^\dag }M$ is the $3 \times 3$ matrix.

For any three-qubit state $\rho_{ABC}$, the analytical trade-off relations of the maximal violation of $\rm {CHSH}$ texts on pairwise bipartite states can be obtained as \cite{qhh}:
\begin{align}
{\left\langle {\rm {CHSH}} \right\rangle ^2}_{{\rho _{AB}}} + {\left\langle {\rm {CHSH}} \right\rangle ^2}_{{\rho _{AC}}} + {\left\langle {\rm {CHSH}} \right\rangle ^2}_{{\rho _{BC}}} \le 12,
\label{Eq.12}
\end{align}
which can be used to study nonlocality distributions of the multi-qubit systems.

The information-theoretic quantification of quantum coherence plays a central role in applications of quantum resource theory.
As an effective coherence measure, the first-order coherence has been widely used in many quantum systems, such as optical systems.
For the three-qubit state $\rho_{ABC}$, the first-order coherence of each subsystem $A$, $B$ or $C$ are
\begin{align}
D({\rho _{A, B, C}}) = \sqrt {2\rm {Tr}(\rho _{A, B, C}^2)- 1},
\label{Eq.18}
\end{align}
where $\rho_A\!=\!$Tr$_{BC}(\rho_{ABC})$, $\rho _B\!=\!$Tr$_{AC}(\rho _{ABC}) $ and $\rho _C\!=\!$Tr$_{AB}(\rho _{ABC})$ are
the reduced density matrix of the composite system ${\rho _{ABC}}$.
The first-order coherence of the composite system is defined as
\begin{align}
D({\rho _{ABC}}) = \sqrt {\frac{{D{{\left( {{\rho _A}} \right)}^2} + D{{\left( {{\rho _B}} \right)}^2} + D{{\left( {{\rho _C}} \right)}^2}}}{3}}.
\label{Eq.17}
\end{align}

On the other hand, for a two-quit pure state $\left| \psi  \right\rangle $, its spin-flipped state is $\left| {\widetilde \psi } \right\rangle=\left( {{\sigma _2} \otimes {\sigma _2}} \right)\left| {{\psi ^ * }} \right\rangle $, where $\left| {{\psi ^*}} \right\rangle$ is the complex conjugate of $\left| \psi  \right\rangle $, ${\sigma _2}$ is the Pauli matrix in the y-direction.
The concurrence, an entanglement measure, is
\begin{align}
C\left( {\left| \psi  \right\rangle } \right) = \left| {\left\langle {\psi |\widetilde \psi } \right\rangle } \right|.
\label{Eq.13}
\end{align}
For an arbitrary two-qubit state $\rho$, its spin-flipped density matrix is $\widetilde {\rho}=({\sigma _2} \otimes {\sigma _2}) {\rho}^\ast ({\sigma _2} \otimes {\sigma _2})$. According to the convex-roof construction \cite{kafv2001}, the concurrence is defined as
\begin{align}
C\left( {{\rho}} \right)=\mathop {\min }\limits_{\{ {P_n},\left| {{\phi _n}} \right\rangle \} } \sum\limits_n {{P_n}} C(\left| {{\phi _n}} \right\rangle ).
\label{Eq.14}
\end{align}
Here, it take over all possible decompositions $\rho$ into pure states by the minimization. By the necessary mathematical calculation and analysis, the concurrence can be simplified as
\begin{align}
C\left( {{\rho}} \right) = \max \left\{ {0,\sqrt {\lambda {}_1}  - \sqrt {{\lambda _2}}  - \sqrt {\lambda {}_3}  - \sqrt {\lambda {}_4} } \right\},
\label{Eq.14}
\end{align}
where ${\lambda _n} (n = 1, 2, 3, 4)$ represent eigenvalues of the non-Hermitian matrix ${\rho}\widetilde {{\rho}}$ and are listed in descending order.

Another proper entanglement measure is the intrinsic concurrence, which is, for the two-qubit state $\rho$, defined as:
\begin{align}
{C_I}({\rho}) = \sqrt {\sum\limits_{n = 1}^4 {{\rm P}_n^2{C^2}(\left| {{\phi _n}} \right\rangle )} },
\label{Eq.first}
\end{align}
where ${\rm P}_n$ is the decomposition probabilities of the pure state ensemble $\rho=\Sigma_{\rm P=1}^4|\phi_n\rangle\langle\phi_n|$.
In general, the relation between concurrence and intrinsic concurrence can be expressed as ${C_I}({\rho}) \geqslant C({\rho})$.
The concurrence and intrinsic concurrence are equivalent as ${C_I}({\rho}) = C({\rho })$ when $P_1 =1$ and $P_2=P_3=P_4=0$, i.e.,
the two-qubit state is pure state. This relation is also effective when the rank of the matrix  $\rho \widetilde {\rho}$ is $0$ or $1$ for
an arbitrary two-qubit state $\rho$ \cite{xgf}. Finally, for the state ${\rho _{ABC}}$, the intrinsic concurrence of the composite system is
\begin{align}
{C_I}({\rho _{ABC}}) = \sqrt {C_I^2({\rho _{AB}}) + C_I^2({\rho _{BC}}) + C_I^2({\rho _{AC}})}.
\label{Eq.15}
\end{align}

For an arbitrary three-qubit pure state, there exists a trade-off relation between the first-order coherence and intrinsic concurrence \cite{fmw}
\begin
{align}{D^2}({\rho _{ABC}}) + \frac{2}{3}C_I^2({\rho _{ABC}}) = 1.
\label{Eq.16}
\end{align}

Then, we discuss the additivity relation of quantum coherence in three-qubit quantum system, based on the relative entropy of coherence. The relative entropy
of coherence is given by
\begin{align}
C(\rho' ) = \mathop {min}\limits_{\delta  \subset {\rm I}} S(\rho' \parallel \delta ) = S({\rho' _{\rm I}}) - S(\rho' ),
\label{Eq.coh}
\end{align}
where $\rm S$ is the von Neumann entropy with
\begin{align}
S({\rho' })=-tr(\rho' \log\rho' ) =  - \sum\limits_j {{\lambda _j}} \log{\lambda _j},
\label{Eq.23}
\end{align}
where $\lambda _j$ corresponds to the eigenvalue of the system state $\rho'$, and ${\rho' _{\rm I}}$ is the diagonal version of $\rho' $, which only retains the diagonal elements of $\rho' $.
The additivity relation of quantum coherence of the tripartite scenario states how quantum coherence is distributed among the subsystems. Explicitly, the tripartite
coherence is equal to or greater than sum of the bipartite coherence, that is \cite{yyx}:
\begin{align}
C({\rho _{ABC}}) \ge C\left( {{\rho _{AC}}} \right) + C\left( {{\rho _{AB}}} \right),
\label{Eq.24}
\end{align}
where $\rho _{AB} = $Tr$_C\left( {{\rho _{ABC}}} \right)$ and $\rho _{AC} = $Tr$_B\left( {{\rho _{ABC}}} \right)$.

\section{Trade-off relation of QRT in NOs}

Here, we shall consider the trade-off relation of QRT in
electron and muon NOs, respectively.

\subsection{Trade-off relation of QRT in the electron antineutrino oscillations}
\begin{figure}
\begin{minipage}{0.45\textwidth}
\centering
\subfigure{\includegraphics[width=8cm]{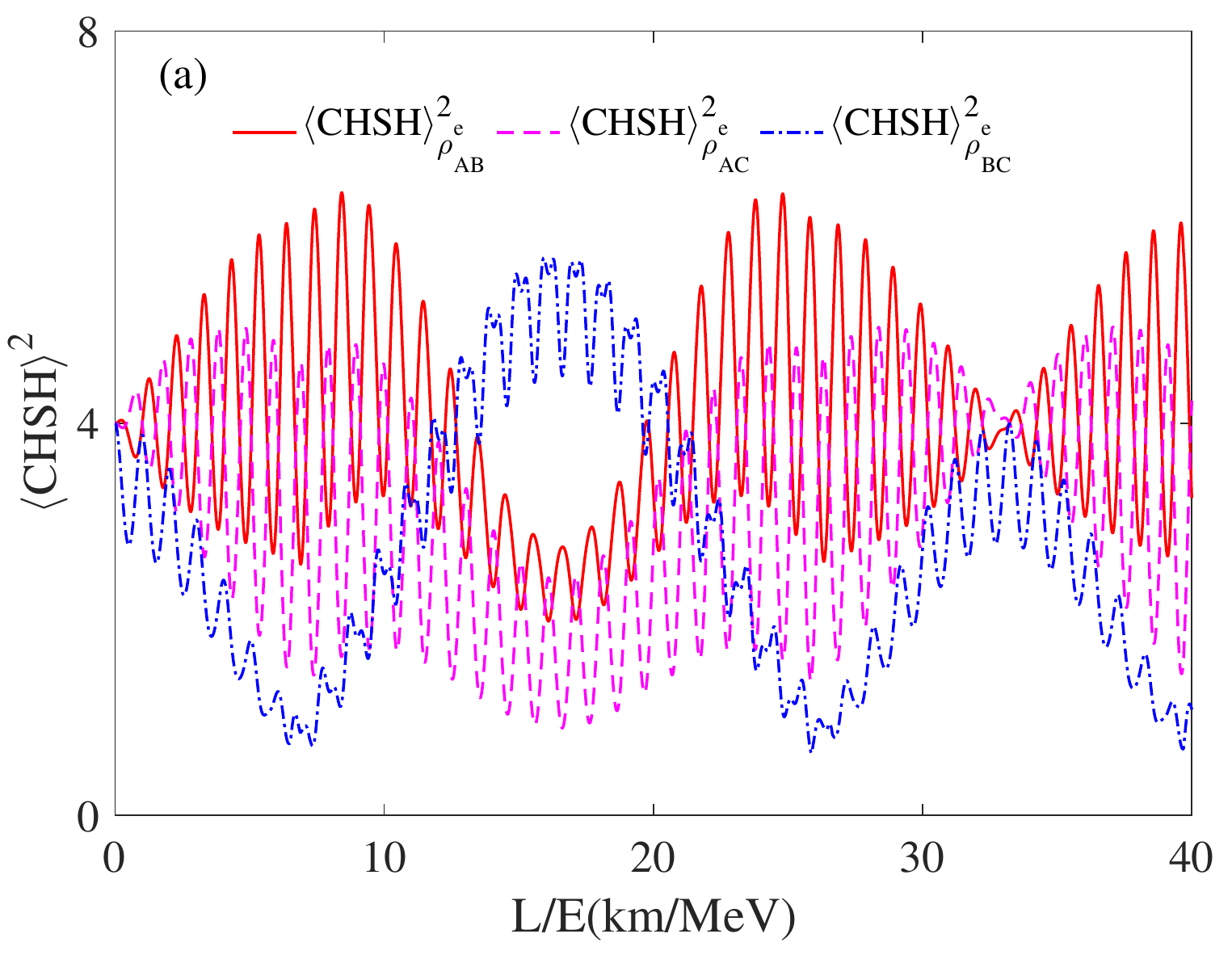}}
\subfigure{\includegraphics[width=8cm]{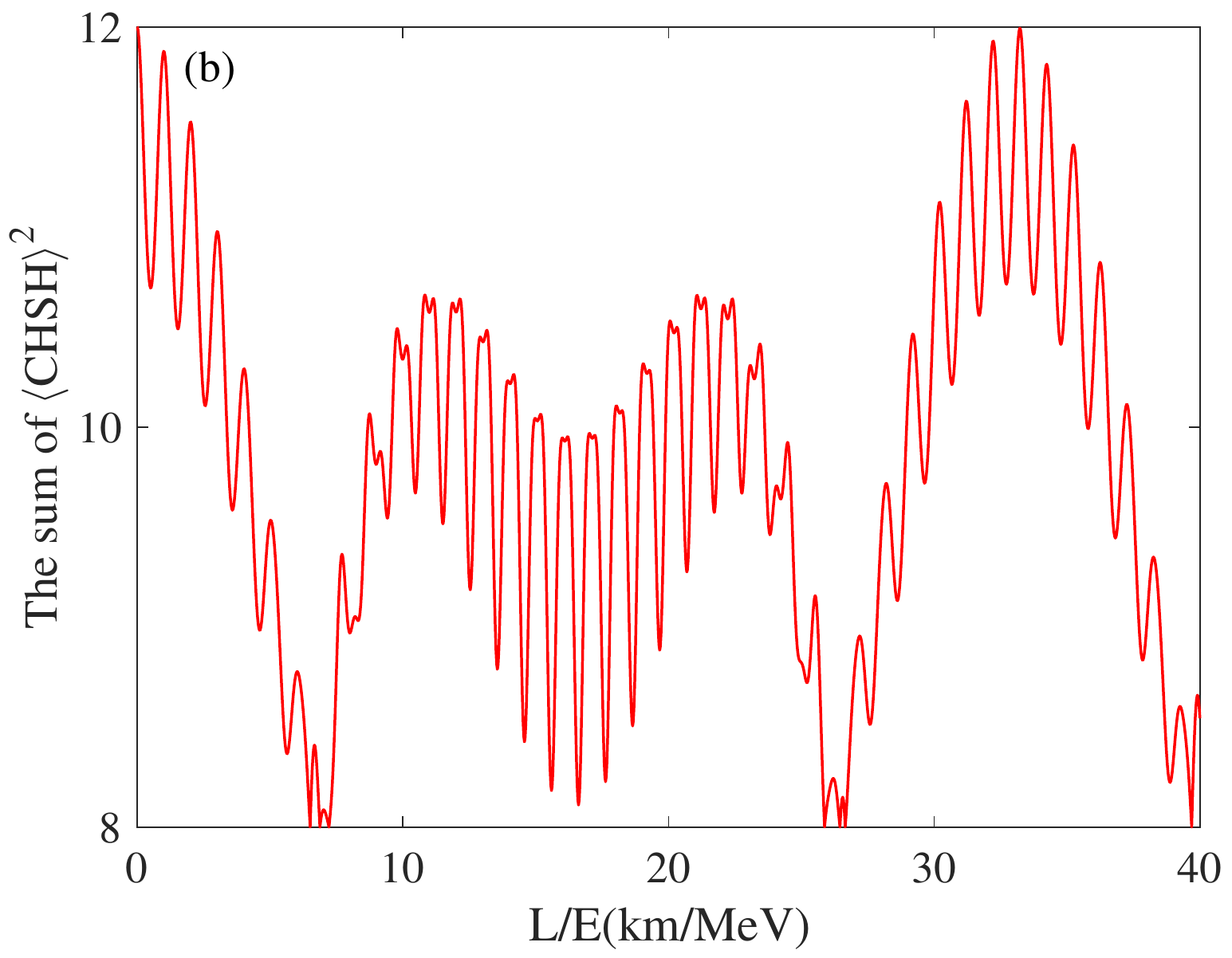}}
\end{minipage}\hfill
\caption{The value of $\rm {CHSH}$ tests for electron. Figure(a) gives $\left\langle {\rm {CHSH}} \right\rangle _{{\rho^e _{AB}}}^2$(red, solid line), $\left\langle {\rm {CHSH}} \right\rangle _{{\rho^e _{AC}}}^2$ (purple, dashed line), $\left\langle {\rm {CHSH}} \right\rangle _{{\rho^e _{BC}}}^2$(blue, dashed-dotted line). Figure(b) is presented as the sum of the  ${\left\langle \rm{CHSH} \right\rangle ^2}_{{\rho^e _{AB}}}$, ${\left\langle \rm{CHSH} \right\rangle ^2}_{{\rho^e _{AC}}}$  and  ${\left\langle \rm{CHSH} \right\rangle ^2}_{{\rho^e _{BC}}}$.}
\label{f2}
\end{figure}

When the electron neutrino in the initial time  $t=0$, the evolutive
states for three-flavor NOs, from Eq. (\ref{Eq.9}), are written as
\begin{align}
{\left| {\psi (t)} \right\rangle _e} = {a_{ee}}(t)\left| {{v_e}} \right\rangle  + {a_{e\mu }}(t)\left| {{v_\mu }} \right\rangle  + {a_{e\tau }}(t)\left| {{v_\tau }} \right\rangle,
\label{Eq.25}
\end{align}
and its density matrix, in the orthonormal basis $\{\left| {000} \right\rangle$, $\left| {001} \right\rangle$, $\left| {010} \right\rangle$, $\left| {011} \right\rangle$, $\left| {100} \right\rangle$, $\left| {101} \right\rangle$,
$\left| {110}\right\rangle$, $\left| {111} \right\rangle\}$, is expressed as $\rho _{ABC}^e(t)={\left| {\psi (t)} \right\rangle _e}\left\langle\psi (t)\right|$. It gives
\begin{eqnarray}
\rho _{ABC}^e(t) = \left( {\begin{array}{*{20}{c}}
0 & 0 & 0 & 0 & 0 & 0 & 0 & 0\\
0 & {\rho _{22}^e} & {\rho _{23}^e} & 0 & {\rho _{25}^e} & 0 & 0 & 0\\
0&{\rho _{32}^e}&{\rho _{33}^e}&0&{\rho _{35}^e}&0&0&0\\
0&0&0&0&0&0&0&0\\
0&{\rho _{52}^e}&{\rho _{53}^e}&0&{\rho _{55}^e}&0&0&0\\
0&0&0&0&0&0&0&0\\
0&0&0&0&0&0&0&0\\
0&0&0&0&0&0&0&0
\end{array}} \right),
\label{Eq.26}
\end{eqnarray}
where the matrix elements are written as\\
$\rho _{22}^e\! =\!{\left| {{a_{e\tau }}(t)} \right|^2};$\,\,$\rho _{23}^e(t)\! =\! {a_{e\tau }}(t)a_{e\mu }^ * (t);$\,\,$\rho _{25}^e(t) \!= \!{a_{e\tau }}(t)a_{ee}^ * (t);$\\
$\rho _{32}^e \!= \!{a_{e\mu }}(t)a_{e\tau }^ * (t);$\,\,$\rho _{33}^e(t) \!= \!{\left| {{a_{e\mu }}(t)} \right|^2};$\,\,$\rho _{35}^e(t) \!= \!{a_{e\mu }}(t)a_{ee}^ * (t);$\\
$\rho _{52}^e \!=\! {a_{ee}}(t)a_{e\tau }^*(t);$\,\,$\rho _{53}^e(t) \!= \!{a_{ee}}(t)a_{e\mu }^*(t);$\,\,$\rho _{55}^e(t)\! = \!{\left| {{a_{ee}}(t)} \right|^2}.$
When traced over the qubit $C$, the reduced density matrix $\rho _{AB}^e$ is given by
\begin{align}
\underline{}\rho _{AB}^e = \rm {Tr}_c(\left| \psi  \right\rangle \left\langle \psi  \right|){\rm{ = }}\left( {\begin{array}{*{20}{c}}
{{{\left| {{a_{e\tau }}} \right|}^2}}&0&0&0\\
0&{{{\left| {{a_{e\mu }}} \right|}^2}}&{{a_{e\mu }}{a_{ee}}^*}&0\\
0&{{a_{ee}}{a_{e\mu }}^*}&{{{\left| {{a_{ee}}} \right|}^2}}&0\\
0&0&0&0
\end{array}} \right).
\label{Eq.27}
\end{align}
So the corresponding correlation matrix M is
\begin{eqnarray}
{M_{AB}^e}=\left( {\begin{array}{*{20}{c}}
{2\sqrt {{P_{e\mu }}{P_{ee}}} }&0&0\\
0&{2\sqrt {{P_{ee}}{P_{e\mu }}} }&0\\
0&0&{{P_{e\tau }} - {P_{e\mu }} - {P_{ee}}}
\end{array}} \right),
\label{Eq.28}
\end{eqnarray}
with the matrix element being $m_{AB({ij})}^e=\rm {Tr}({\rho _{AB}^e}{\sigma _i} \otimes {\sigma _j})$. Similarly, we can get the correlation matrices
${M_{AC}^e}$ and ${M_{BC}^e}$. Then the square of the CHSH test of pairwise qubits in the three-flavor electron neutrino system are calculated as
\begin{align}
{\left\langle \rm{CHSH} \right\rangle ^2}_{{\rho^e _{AB}}} = 2\left\{ {4{P_{ee}}{P_{e\mu }} + max\left[ {4{P_{ee}}{P_{e\mu }},{{(2{P_{e\tau }} - 1)}^2}} \right]} \right\},\nonumber \\
{\left\langle \rm{CHSH} \right\rangle ^2}_{{\rho^e _{AC}}} = 2\left\{ {4{P_{ee}}{P_{e\tau }} + max\left[ {4{P_{ee}}{P_{e\tau }},{{(2{P_{e\mu }} - 1)}^2}} \right]} \right\},\nonumber \\
{\left\langle \rm{CHSH} \right\rangle ^2}_{{\rho^e _{BC}}} = 2\left\{ {4{P_{e\mu }}{P_{e\tau }} + max\left[ {4{P_{e\mu }}{P_{e\tau }},{{(2{P_{ee}} - 1)}^2}} \right]} \right\},
\label{Eq.29}
\end{align}
which are plotted in Fig. \ref{f2} (a). Also, to study the trade-off relations of Bell violations among pairwise neutrino flavor systems, we plot the sum of ${\left\langle \rm{CHSH} \right\rangle ^2}_{{\rho^e _{AB}}}$, ${\left\langle \rm{CHSH} \right\rangle ^2}_{{\rho^e _{AC}}}$, and ${\left\langle \rm{CHSH} \right\rangle ^2}_{{\rho^e _{BC}}}$ in Fig. \ref{f2} (b). One can see that most of the CHSH value of subsystem $AB$ is larger than $2$, meaning the violation of Bell inequality. The monogamy relation ${\left\langle \rm{CHSH} \right\rangle ^2}_{{\rho^e _{AB}}}$
+${\left\langle \rm{CHSH} \right\rangle ^2}_{{\rho^e _{AC}}}$
+ ${\left\langle \rm{CHSH} \right\rangle ^2}_{{\rho^e _{BC}}}\leq12$ is valid in the whole traveled distance of $L/E$ for the electron antineutrino oscillations.
This trade-off relation shows a strong restriction on the maximal violations
among reduced two-flavor neutrino systems.

\begin{figure}
\centering
\includegraphics[width=8cm]{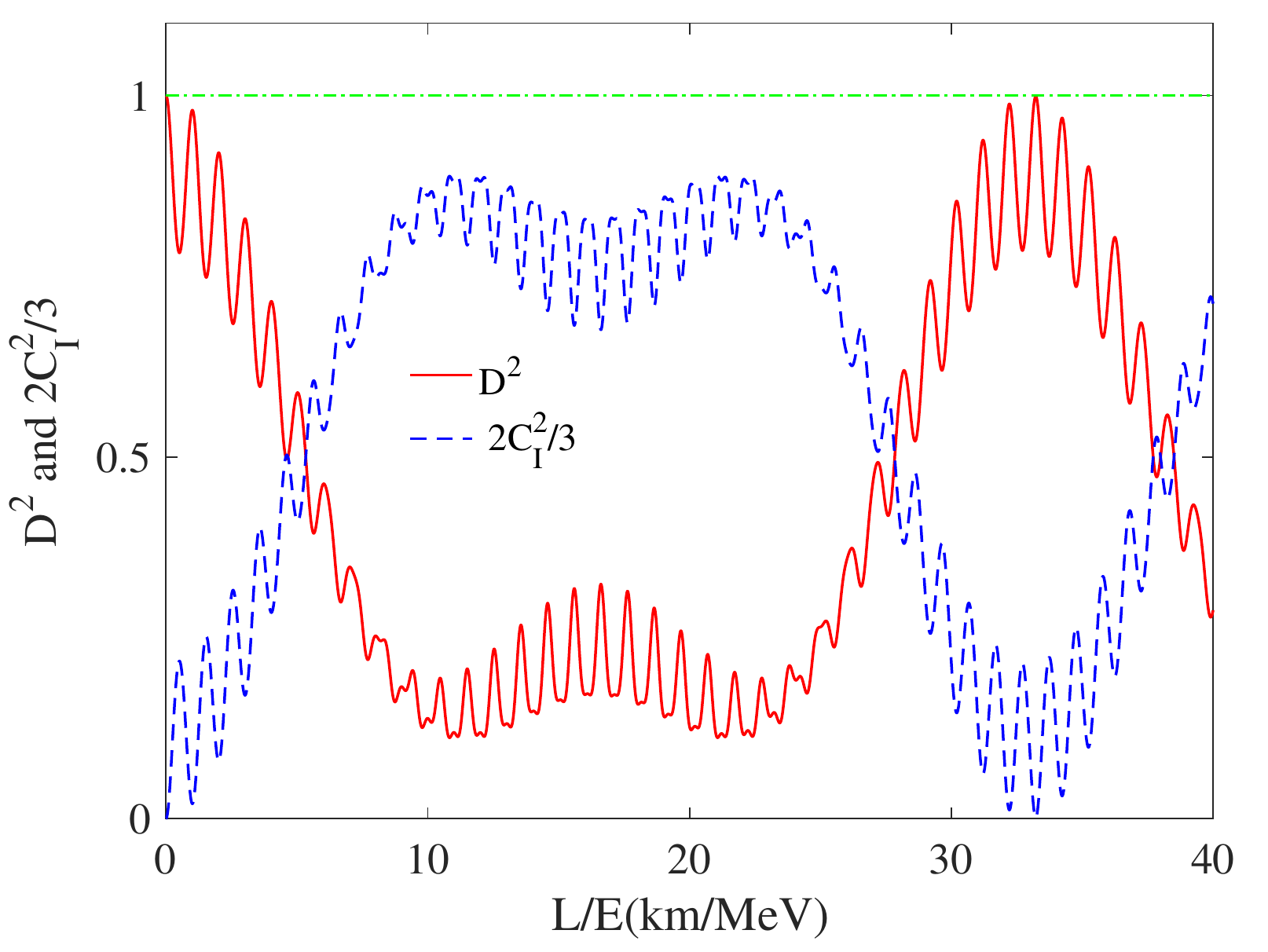}
\caption{${D^2}({\rho^e _{ABC}})$ (red, solid line) and $2C_I^2({\rho^e _{ABC}})/3$ (blue, dashed line) in the electron
antineutrino oscillations. One can see that it satisfies the relation ${D^2}({\rho^e _{ABC}}) + 2C_I^2({\rho^e _{ABC}})/3 = 1$. }
\label{f3}
\end{figure}

To obtain the first-order coherence of the three-qubit neutrino system, we should first get the reduced density matrix of the composite system ${\rho^e _{ABC}}$: ${\rho^e _A}$, ${\rho^e _B} $, and ${\rho^e _C} $, by
tracing over the corresponding subsystems. After the calculation, we have: ${\rho^e _A} = diag\{ {P_{e\mu }} + {P_{e\tau }}, {P_{ee}}\} $, ${\rho^e _B} = diag\{ {P_{e\tau }} + {P_{ee}}, {P_{e\mu }}\} $, and ${\rho^e _C} = diag\{ {P_{ee}} + {P_{e\mu }}, {P_{e\tau }}\} $.
Then the first-order coherence of each subsystem $A$, $B$ or $C$, from Eq. (\ref{Eq.18}), are given by
\begin{align}
&D({\rho^e _A}) = \sqrt {2(P_{ee}^2 + P_{e\mu }^2 + P_{e\tau }^2 + 2{P_{e\tau }}{P_{e\mu }}) - 1},\nonumber \\
&D({\rho^e _B}) = \sqrt {2(P_{ee}^2 + P_{e\mu }^2 + P_{e\tau }^2 + 2{P_{e\tau }}{P_{ee}}) - 1},\nonumber \\
&D({\rho^e _C}) = \sqrt {2(P_{ee}^2 + P_{e\mu }^2 + P_{e\tau }^2 + 2{P_{ee}}{P_{e\mu }}) - 1}.
\label{Eq.30}
\end{align}
Finally, the first-order coherence of the three-flavor electron antineutrino oscillations is obtained as:
\begin{align}
{D}({\rho^e _{ABC}})\!\!\!=\!\!\!\sqrt{2(P_{ee}^2\!+\!P_{e\mu }^2\!+\!P_{e\tau }^2)\!+\!\frac{4}{3}({P_{ee}}{P_{e\mu }}\!\!+\!\!{P_{ee}}{P_{e\tau }}\!+\!{P_{e\mu }}{P_{e\tau }})\!-\!1}.
\label{Eq.31}
\end{align}
For the reduced density matrix state $\rho _{AB}^e$ in  the three-flavor electron antineutrino oscillations, the matrix $\rho _{AB}^e\widetilde {\rho _{AB}^e}$ can be expressed as
\begin{align}
{\rho _{AB}^e}\widetilde {\rho _{AB}^e}= \left( {\begin{array}{*{20}{c}}
0&0&0&0\\
0&{2{P_{ee}}{P_{e\mu }}}&{2{P_{e\mu }}\sqrt {{P_{ee}}{P_{e\mu }}} }&0\\
0&{2{P_{ee}}\sqrt {{P_{ee}}{P_{e\mu }}} }&{2{P_{ee}}{P_{e\mu }}}&0\\
0&0&0&0
\end{array}} \right)
\label{Eq.32}.
\end{align}
Its has a non-zero eigenvalue, and the concurrence is equal to the intrinsic concurrence. The intrinsic concurrence of the reduced matrix density
 $\rho _{AB}^e$, $\rho _{AC}^e$, and $\rho _{BC}^e$ can be calculated as
\begin{align}
&{C_I}({\rho^e _{AB}}) = \sqrt {4{P_{ee}}{P_{e\mu }}},\\
&{C_I}({\rho^e _{AC}}) = \sqrt {4{P_{ee}}{P_{e\tau }}},\\
&{C_I}({\rho^e _{BC}}) = \sqrt {4{P_{e\mu }}{P_{e\tau }}}
\label{Eq.33}.
\end{align}
The intrinsic concurrence of ${\rho^e _{ABC}}$, using Eq. (\ref{Eq.15}), is
\begin{align}
C_I({\rho^e _{ABC}}) =2 \sqrt{({P_{ee}}{P_{e\mu }} + {P_{ee}}{P_{e\tau }} + {P_{e\mu }}{P_{e\tau }})}.
\label{Eq.34}
\end{align}
Finally, using Eqs. (\ref{Eq.31}) and (\ref{Eq.34}), we obtain the relation that
\begin{align}
{D^2}({\rho^e _{ABC}}) + \frac{2}{3}C_I^2({\rho^e _{ABC}}) = 1,
\label{Eq.35}
\end{align}
which shows a trade-off relation between first-order coherence and intrinsic concurrence for the three-flavor neutrino quantum system.
In Fig. \ref{f3} , we plot how the $2C_I^2({\rho^e _{ABC}})/3$ and ${D^2}({\rho^e _{ABC}})$ change with the variation of the ratio $L/E$. The
${D}({\rho^e _{ABC}})$ decreases firstly and then increases, and $C_I({\rho^e _{ABC}})$ increases firstly and then decreases. At the point $L/E=0$,
$2C_I^2({\rho^e _{ABC}})/3=0$ and ${D^2}({\rho^e _{ABC}})=1$. When the $2C_I^2({\rho^e _{ABC}})/3$ increases the maximal point about $0.89$ at around
$L/E = 10.8~\rm km/MeV$, and $D^2({\rho^e _{ABC}})$ reaches its minimal value with $0.11$. Furthermore, we can find that the sum of
${D^2}({\rho^e _{ABC}})$ and $2C_I^2({\rho^e _{ABC}})/3$ always equals to $1$ with respect to $L/E$.

Now, we study the additivity relation of quantum coherence in the electron antineutrino
oscillation system. According to Eq. (\ref{Eq.coh}), we obtain the relative entropy of coherence for the subsystem $\rho _{AB}^e$, $\rho _{AC}^e$, and $\rho _{ABC}^e$ as
\begin{align}
{C\left( {{\rho^e _{AC}}} \right) =  - {P_{ee}}{{\log }_2}\frac{{{P_{ee}}}}{{{P_{ee}} + {P_{e\tau }}}} + {P_{e\tau }}{{\log }_2}\frac{{{P_{e\tau }}}}{{{P_{ee}} + {P_{e\tau }}}}},
\label{Eq.21}
\end{align}
\begin{align}
{C\left( {{\rho^e _{AB}}} \right) =  - {P_{e\mu }}{{\log }_2}\frac{{{P_{e\mu }}}}{{{P_{ee}} + {P_{e\mu }}}} + {P_{ee}}\log_2 \frac{{{P_{ee}}}}{{{P_{ee}} + {P_{e\mu }}}}},
\label{Eq.21}
\end{align}
\begin{align}
{C\left( {{\rho^e _{ABC}}} \right) =  - {P_{ee}}{{\log }_2}{P_{ee}} - {P_{e\mu }}{{\log }_2}{P_{e\mu }} - {P_{e\tau }}{{\log }_2}{P_{e\tau }}},
\label{Eq.21}
\end{align}
respectively. Thus, we have the inequality
\begin{eqnarray}
Q_{\rho^e _{ABC}}
\! &=&\! -{P_{ee}}\log_2{P_{ee}}\!+\!(1\! -\! {P_{e\mu }})\log_2(1 \!-\! {P_{e\mu }})\nonumber\\&&+ (1 - {P_{e\tau }})\log_2(1 - {P_{e\tau }})\leq0,
\label{Eq.ABCE}
\end{eqnarray}
where $Q_{\rho^e _{ABC}}=C\left( {{\rho^e _{AC}}} \right) + C\left( {{\rho^e _{AB}}} \right) - C({\rho^e _{ABC}})$. The last inequality holds because for given $P_{ee}$,
assuming that $x=1-{P_{e\mu } }\leq 1$ and $a=1+P_{ee}\geq1$, the function $G(x)=x\log_2 {x}+(a-x)\log_2 {(a-x)}$ is a concave function in  $x\leq a$ by taking the second derivative. Thus, the maximum value of $G(x)$ is reached at the boundary, $P_{e\mu}=0$ or $P_{e\tau}=0$.

\begin{figure}
\centering
\includegraphics[width=8cm]{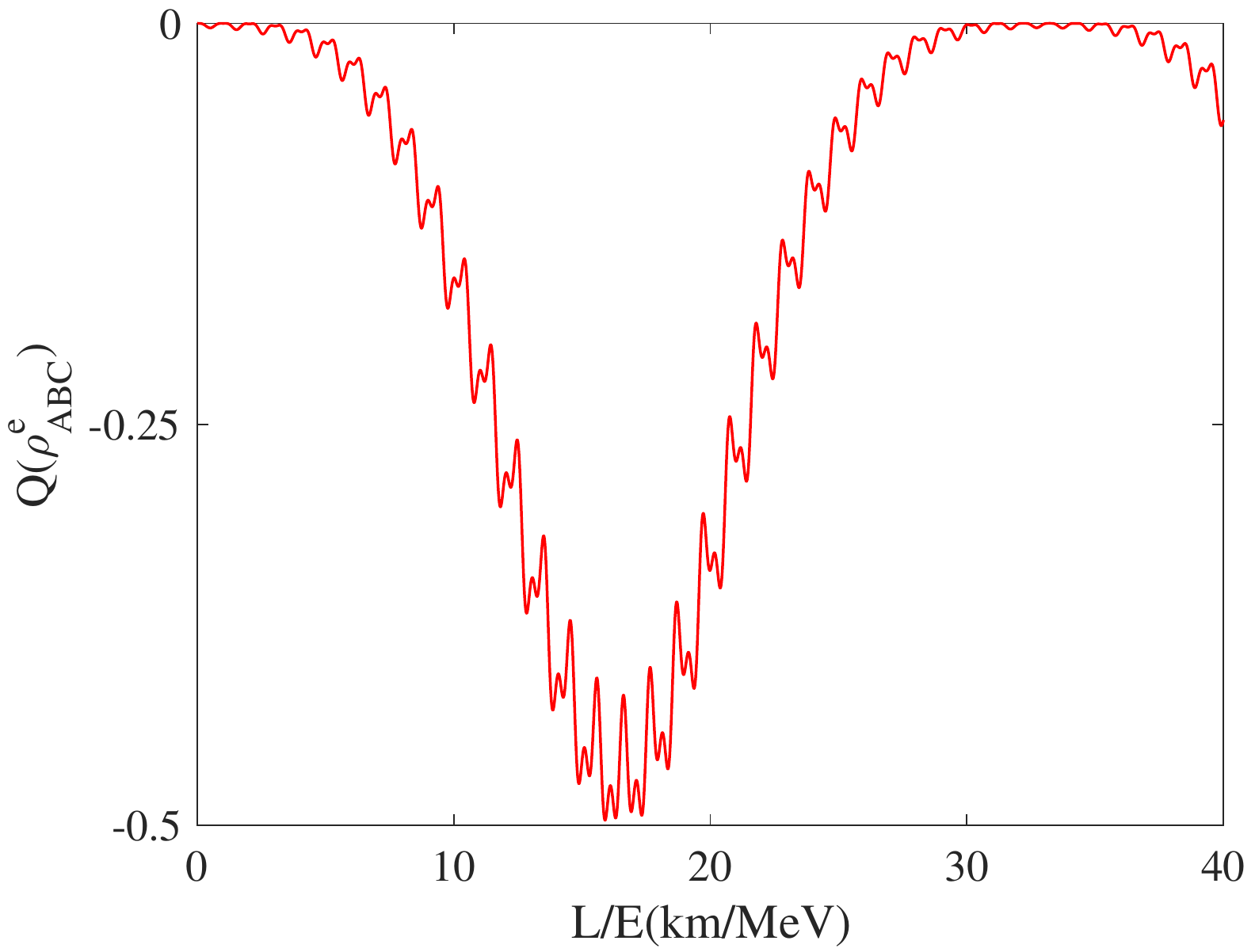}
\caption{The additivity relation $\left[C\left( {{\rho^\mu _{AC}}} \right) + C\left( {{\rho^\mu _{AB}}} \right) - C({\rho^\mu _{ABC}})\right]$ (red, solid line) of quantum coherence in the electron antineutrino oscillation system. }
\label{f4}
\end{figure}

In Fig. \ref{f4} , we plot the additivity relation of the relative entropy of coherence for the electron antineutrino oscillation system. It can be seen that
$Q_{\rho^e _{ABC}}$ is always less than or equal to $0$, implying that the inequality in Eq. (\ref{Eq.ABCE}) holds in this three-flavor neutrino oscillations. This suggests
that the tripartite coherence of three-flavor NOs contains coherence of the subsystems between $A$, $B$ and $B$, $C$, but they are not exactly
same.

\subsection{Trade-off relation of QRT in the muon antineutrino oscillations}

\begin{figure}
\begin{minipage}{0.45\textwidth}
\centering
\subfigure{\includegraphics[width=8cm]{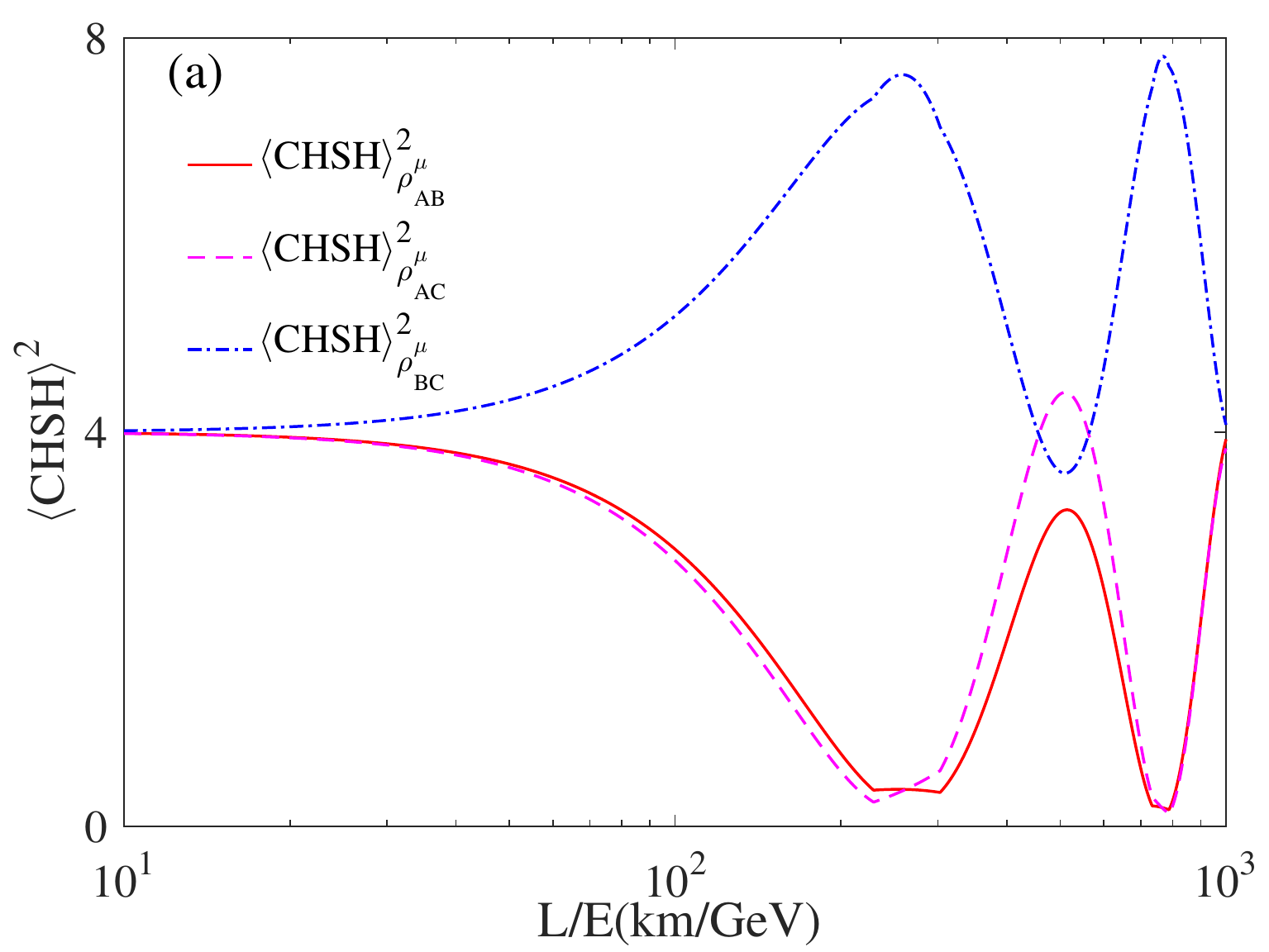}}
\subfigure{\includegraphics[width=8cm]{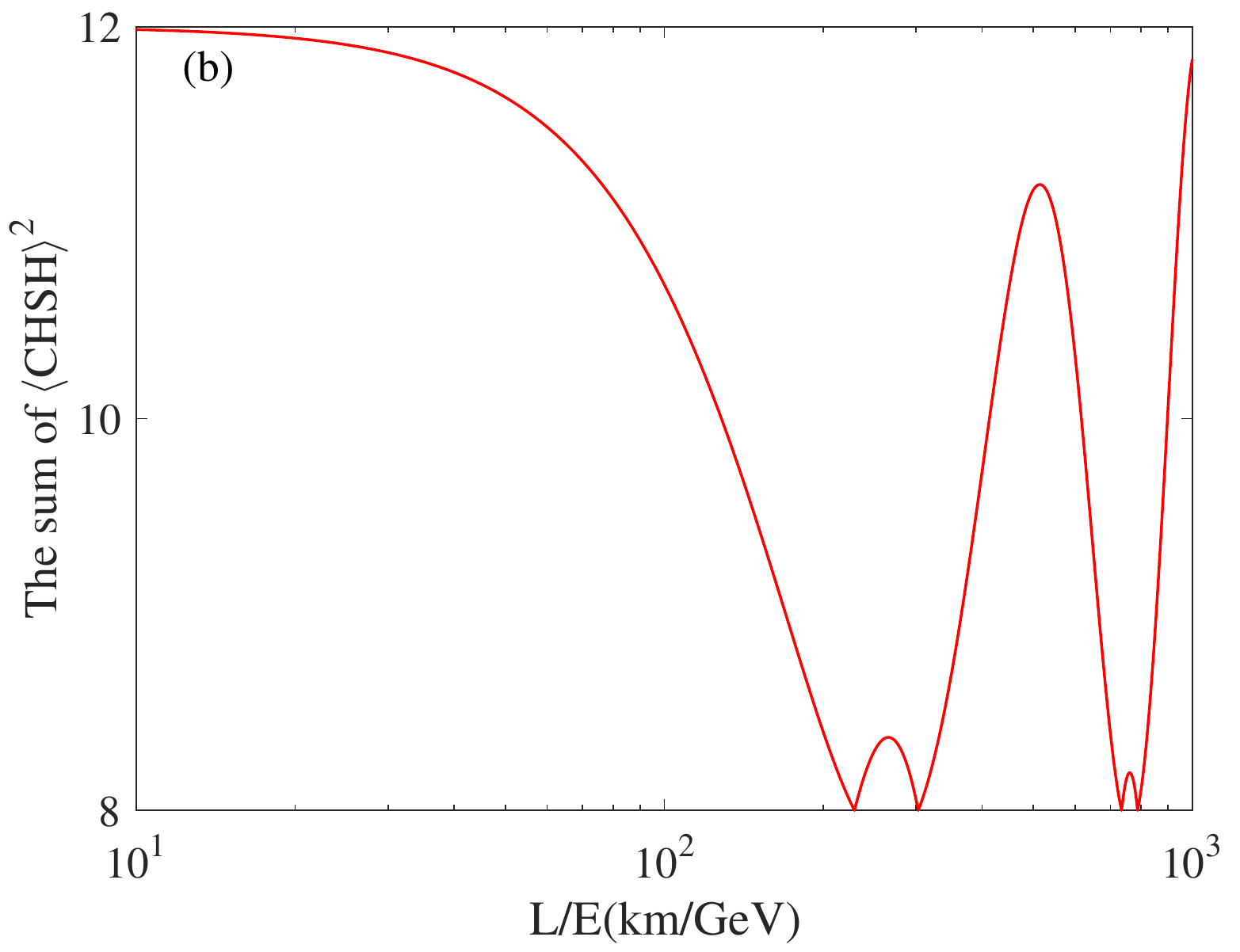}}
\end{minipage}\hfill
\caption{The maximal value of $\rm {CHSH}$ tests for muon neutrino oscillations. Figure(a) gives $\left\langle {\rm {CHSH}} \right\rangle _{{\rho^\mu _{AB}}}^2$(red, solid line), $\left\langle {\rm {CHSH}} \right\rangle _{{\rho^\mu  _{AC}}}^2$ (purple, dashed line), $\left\langle {\rm {CHSH}} \right\rangle _{{\rho^\mu  _{BC}}}^2$(blue, dashed-dotted line). Figure(b) is presented as the sum of the ${\left\langle \rm{CHSH} \right\rangle ^2}_{{\rho^\mu _{AB}}}$, ${\left\langle \rm{CHSH} \right\rangle ^2}_{{\rho^\mu _{AC}}}$  and  ${\left\langle \rm{CHSH} \right\rangle ^2}_{{\rho^\mu _{BC}}}$. (red, solid line).}
\label{f5}
\end{figure}

We begin by preparing a muon neutrino in the initial time $t=0$, from Eq. (\ref{Eq.9}), the evolution of states for the three-flavor NOs is
\begin{align}
{\left| {\psi (t)} \right\rangle _\mu } = {a_{\mu e}}(t)\left| {{v_e}} \right\rangle  + {a_{\mu \mu }}(t)\left| {{v_\mu }} \right\rangle  + {a_{\mu \tau }}(t)\left| {{v_\tau }} \right\rangle,
\label{Eq.50}
\end{align}
and its density matrix, in the orthonormal basis $\{\left| {000} \right\rangle$, $\left| {001} \right\rangle$, $\left| {010} \right\rangle$, $\left| {011} \right\rangle$, $\left| {100} \right\rangle$, $\left| {101} \right\rangle$,
$\left| {110}\right\rangle$, $\left| {111} \right\rangle\}$, is expressed as $\rho _{ABC}^\mu(t)={\left| {\psi (t)} \right\rangle _\mu}\left\langle\psi (t)\right|$. This gives
\begin{eqnarray}
\rho _{ABC}^\mu (t) = \left( {\begin{array}{*{20}{c}}
0&0&0&0&0&0&0&0\\
0&{\rho _{22}^\mu}&{\rho _{23}^\mu}&0&{\rho _{25}^\mu }&0&0&0\\
0&{\rho _{32}^\mu }&{\rho _{33}^\mu }&0&{\rho _{35}^\mu }&0&0&0\\
0&0&0&0&0&0&0&0\\
0&{\rho _{52}^\mu}&{\rho _{53}^\mu}&0&{\rho _{55}^\mu}&0&0&0\\
0&0&0&0&0&0&0&0\\
0&0&0&0&0&0&0&0\\
0&0&0&0&0&0&0&0
\end{array}} \right),
\label{Eq.38}
\end{eqnarray}
where the matrix elements are given by\\
$\rho _{22}^\mu\!=\! {\left| {{a_{\mu \tau }}(t)} \right|^2}$; \,\,$\rho _{23}^\mu (t) \!=\! {a_{\mu \tau }}(t)a_{\mu \mu }^ * (t)$; \,\,$\rho _{25}^\mu\!=\!{a_{\mu \tau }}(t)a_{\mu e}^ * (t);$\\
$\rho _{32}^\mu\!\!=\!{a_{\mu \mu }}(t)a_{\mu \tau }^ * (t)$; \,\,$\rho _{33}^\mu (t)\!\!=  \!\!{\left| {{a_{\mu \mu }}(t)} \right|^2}$; \,\,$\rho _{35}^\mu (t)\!=\!{a_{\mu \mu }}(t)a_{\mu e}^ * (t);$\\
$\rho _{52}^\mu\!=\!{a_{\mu e}}(t)a_{\mu \tau }^*(t)$; \,\,$\rho _{53}^\mu (t)\! =\! {a_{\mu e}}(t)a_{\mu \mu }^*(t)$; \,\,$\rho _{55}^\mu (t)\!=\!{\left|{{a_{\mu e}}(t)}\right|^2}.$
The reduced density matrix $\rho _{AB}^\mu$ is given by tracing over the qubit $C$
\begin{align}
{\rho _{AB}^\mu}\! =\!\rm {Tr}_c(\left| \psi  \right\rangle \langle \psi |){\rm{ = }}\left( {\begin{array}{*{20}{c}}
{{{\left| {{a_{\mu \tau }}} \right|}^2}}&0&0&0\\
0&{{{\left| {{a_{\mu \mu }}} \right|}^2}}&{{a_{\mu \mu }}{a_{\mu e}}^*}&0\\
0&{{a_{\mu e}}{a_{\mu \mu }}^*}&{{{\left| {{a_{\mu e}}} \right|}^2}}&0\\
0&0&0&0
\end{array}} \right).
\label{Eq.39}
\end{align}
The corresponding correlation matrices $M$ for subsystems $\rho _{AB}^\mu$, $\rho _{AC}^\mu$, and $\rho _{BC}^\mu$ can be easily obtained.
Then the square of the CHSH test of pairwise qubits are calculated as
\begin{align}
&{\left\langle {\rm CHSH} \right\rangle ^2}_{{\rho^\mu _{AB}}}=2\left\{ {4{P_{\mu e}}{P_{\mu \mu }}
+max\left[ {4{P_{\mu e}}{P_{\mu \mu }},{{(2{P_{\mu \tau }} - 1)}^2}} \right]} \right\},\nonumber \\
&{\left\langle {\rm CHSH} \right\rangle ^2}_{{\rho^\mu _{AC}}} =2\left\{ {4{P_{\mu e}}{P_{\mu \tau }}
+max\left[ {4{P_{\mu e}}{P_{\mu \tau }},{{(2{P_{\mu \mu }} - 1)}^2}} \right]} \right\},\nonumber \\
&{\left\langle {\rm CHSH} \right\rangle ^2}_{{\rho^\mu _{BC}}}=2\left\{ {4{P_{\mu \mu }}{P_{\mu \tau }}
+max\left[ {4{P_{\mu \mu }}{P_{\mu \tau }},{{(2{P_{\mu e}} - 1)}^2}} \right]} \right\}.
\label{Eq.41}
\end{align}
They are plotted in Fig. \ref{f5} (a), together with their sum in Fig. \ref{f5} (b). While the subsystem $\rho _{BC}^\mu$ violates the Bell-CHSH inequality,
${\left\langle \rm{CHSH} \right\rangle ^2}_{{\rho^\mu _{BC}}}\geq4$, in the most of the ratio $L/E$ range, the other two obey the inequality.
In particular, when $\rho _{BC}^\mu$ is extremely close to the maximal violation of the CHSH inequality, the other two pairs of qubits can no longer violate the
CHSH inequality. That is, ${\left\langle \rm{CHSH} \right\rangle ^2}_{{\rho^\mu _{AB}}}$
+${\left\langle \rm{CHSH} \right\rangle ^2}_{{\rho^\mu _{AC}}}\leq4$,
which implies that ${\left\langle \rm{CHSH} \right\rangle }_{{\rho^\mu _{AC}}}\leq2$ and
${\left\langle \rm{CHSH} \right\rangle}_{{\rho^\mu _{AB}}}\leq2$.
Furthermore, the relation ${\left\langle \rm{CHSH} \right\rangle ^2}_{{\rho^\mu _{AB}}}
+{\left\langle \rm{CHSH} \right\rangle ^2}_{{\rho^\mu _{AC}}}+{\left\langle \rm{CHSH} \right\rangle ^2}_{{\rho^\mu _{BC}}}\leq12$ holds in a range
$[10^1,10^3]$ of $L/E$ with dimension $\rm km/GeV$ for the muon antineutrino oscillations.
This trade-off relation shows importance nonlocality distributions among the three-flavor neutrino systems.

To detect the first-order coherence of the three-qubit neutrino system, we can resort to Eq. (\ref{Eq.18}). At first, we can get the reduced density matrixs of the composite system ${\rho^\mu _{ABC}}$, which are expressed as:
${\rho^\mu _A} = diag\{ {P_{\mu \tau }}+{P_{\mu \mu }}, {P_{\mu e}}\}$, ${\rho^\mu _B} = diag\{ {P_{\mu \tau }} + {P_{\mu e}}, {P_{\mu \mu }}\}$ and ${\rho^\mu _C} = diag\{ {P_{\mu e}} + {P_{\mu \mu }}, {P_{\mu \tau }}\}$, respectively.
The first-order coherence of each subsystem $A$, $B$ or $C$ are given by
\begin{align}
&D({\rho^\mu _A}) = \sqrt {2(P_{\mu e}^2 + P_{\mu \mu }^2 + P_{\mu \tau }^2 + 2{P_{\mu \tau }}{P_{\mu \mu }}) - 1} ,\nonumber \\
&D({\rho^\mu _B}) = \sqrt {2(P_{\mu e}^2 + P_{\mu \mu }^2 + P_{\mu \tau }^2 + 2{P_{\mu \tau }}{P_{\mu e}}) - 1} ,\nonumber \\
&D({\rho^\mu _C}) = \sqrt {2(P_{\mu e}^2 + P_{\mu \mu }^2 + P_{\mu \tau }^2 + 2{P_{\mu e}}{P_{\mu \mu }}) - 1}.
\label{Eq.42}
\end{align}
Then, the first-order coherence of the three-flavor muon antineutrino oscillations, using Eq. (\ref{Eq.17}), can be expressed as:
\begin{align}
D({\rho^\mu _{ABC}})\!\!\!=\!\!\!\sqrt{2(P_{\mu e}^2\!\!+\!\!P_{\mu \mu }^2\!\!+\!\!P_{\mu \tau }^2)\!\!+\!\!\frac{3}{4}({P_{\mu e}}{P_{\mu \mu }}\!\!+\!\!{P_{\mu e}}{P_{\mu \tau }}\!\!+\!\!{P_{\mu \mu }}{P_{\mu \tau }})\!\!-\!\!1}
\label{Eq.43}.
\end{align}
Based on the reduced density matrix state $\rho _{AB}^\mu$, ${\rho^\mu _{AC}}$, ${\rho^\mu _{BC}}$, from Eq. (\ref{Eq.first}), we can get their intrinsic concurrences, respectively.
Then, using Eq. (\ref{Eq.15}), the intrinsic concurrence of the whole three-flavor muon antineutrino oscillations with ${\rho^\mu _{ABC}}$ is
\begin{align}
C_I({\rho^\mu _{ABC}}) = 2\sqrt{({P_{\mu e}}{P_{\mu \mu }} + {P_{\mu e}}{P_{\mu \tau }} + {P_{\mu \mu }}{P_{\mu \tau }})}.
\label{Eq.46}
\end{align}

Using Eqs. (\ref{Eq.31}) and (\ref{Eq.34}), we can obtain the relation that
\begin{align}
{D^2}({\rho^\mu _{ABC}}) + 2C_I^2({\rho^\mu _{ABC}})/3 = 1.
\label{Eq.35}
\end{align}

\begin{figure}
\centering
\includegraphics[width=8cm]{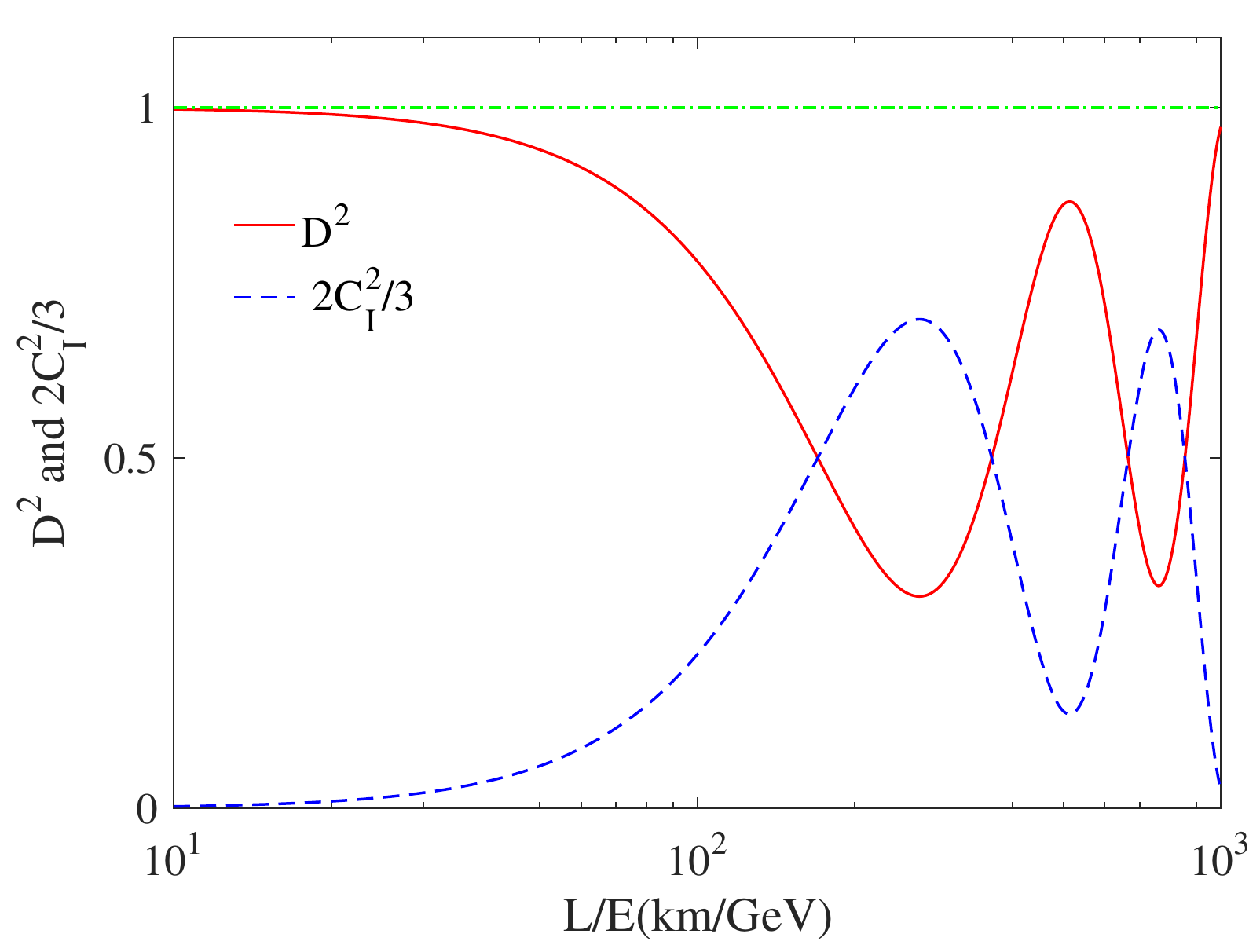}
\caption{${D^2}({\rho^\mu _{ABC}})$ (red, solid line) and $2C_I^2({\rho^\mu _{ABC}})/3$ (blue, dashed line) in the muon
antineutrino oscillations. It satisfies the trade-off relation
${D^2}({\rho^\mu _{ABC}}) + 2C_I^2({\rho^\mu _{ABC}})/3 = 1$.}
\label{f6}
\end{figure}

In Fig. \ref{f6} , we plot how the variation of $2C_I^2({\rho^\mu _{ABC}})/3$ and ${D^2}({\rho _{ABC}})$ with the increase of $L/E$. While the
$2C_I^2({\rho^\mu _{ABC}})/3$ always varies from $0$ to $0.7$, and ${D^2}({\rho _{ABC}})$ changes in a range $[0.3, 1]$. Particularly, at around $L/E = 264.9 ~ \rm km/GeV$,
 $2C_I^2({\rho^\mu _{ABC}})/3$ increases the maximal value with $0.7$ , when $D^2({\rho^\mu _{ABC}})$ reaches its minimal value with $0.3$. Moreover, there exists a close relation between the
$2C_I^2({\rho^\mu _{ABC}})/3$ and ${D^2}({\rho _{ABC}})$, namely, one must decrease as the other increases. But, the sum of
${D^2}({\rho _{ABC}})$ and $2C_I^2({\rho^\mu _{ABC}})/3$ always equals to $1$ with respect to the ratio $L/E$.

Now, we want to know how the correlations in the total neutrino system are distributed among the distinct subsystems, by studying the additivity relation of quantum coherence in the three-flavor muon NOs. According to Eq. (\ref{Eq.coh}), the relative entropy of coherence for the subsystem $\rho _{AB}^\mu$, $\rho _{AC}^\mu$, and $\rho _{ABC}^\mu$ are given by
\begin{align}
{C\left( {{\rho^\mu _{AC}}} \right) =  - {P_{\mu e}}{{\log }_2}\frac{{{P_{ee}}}}{{{P_{\mu e}} + {P_{\mu \tau }}}} + {P_{\mu \tau }}{{\log }_2}\frac{{{P_{\mu \tau }}}}{{{P_{ee}} + {P_{\mu \tau }}}}},
\label{Eq.21}
\end{align}
\begin{align}
{C\left( {{\rho^\mu _{AB}}} \right) =  - {P_{\mu \mu }}{{\log }_2}\frac{{{P_{\mu \mu }}}}{{{P_{\mu e}} + {P_{\mu \mu }}}} + {P_{\mu e}}\log_2 \frac{{{P_{\mu e}}}}{{{P_{\mu e}} + {P_{\mu \mu }}}}},
\label{Eq.21}
\end{align}
\begin{align}
{C\left( {{\rho^\mu _{ABC}}} \right)\! = \!  - {P_{\mu e}}{{\log }_2}{P_{\mu e}} - {P_{\mu \mu }}{{\log }_2}{P_{\mu \mu }} - {P_{\mu \tau }}{{\log }_2}{P_{\mu \tau }}},
\label{Eq.21}
\end{align}
respectively.
Therefore, the inequality of additivity relation of quantum coherence in this system are obtained as
\begin{eqnarray}
Q_{\rho^\mu _{ABC}} &=&-{P_{\mu e}}\log_2{P_{\mu e}}+(1 - {P_{\mu \mu }})\log_2(1 - {P_{\mu \mu }}) \\&&+ (1 - {P_{\mu \tau }})\log_2(1 - {P_{\mu \tau }})
 \leq0,
\label{Eq.48}
\end{eqnarray}
where $Q_{\rho^\mu _{ABC}}=C\left( {{\rho^\mu _{AC}}} \right) + C\left( {{\rho^\mu _{AB}}} \right) - C({\rho^\mu _{ABC}})$. Also, we can prove that $Q_{\rho^\mu _{ABC}}$ must be less than or equal to $0$.

\begin{figure}
\centering
\includegraphics[width=8cm]{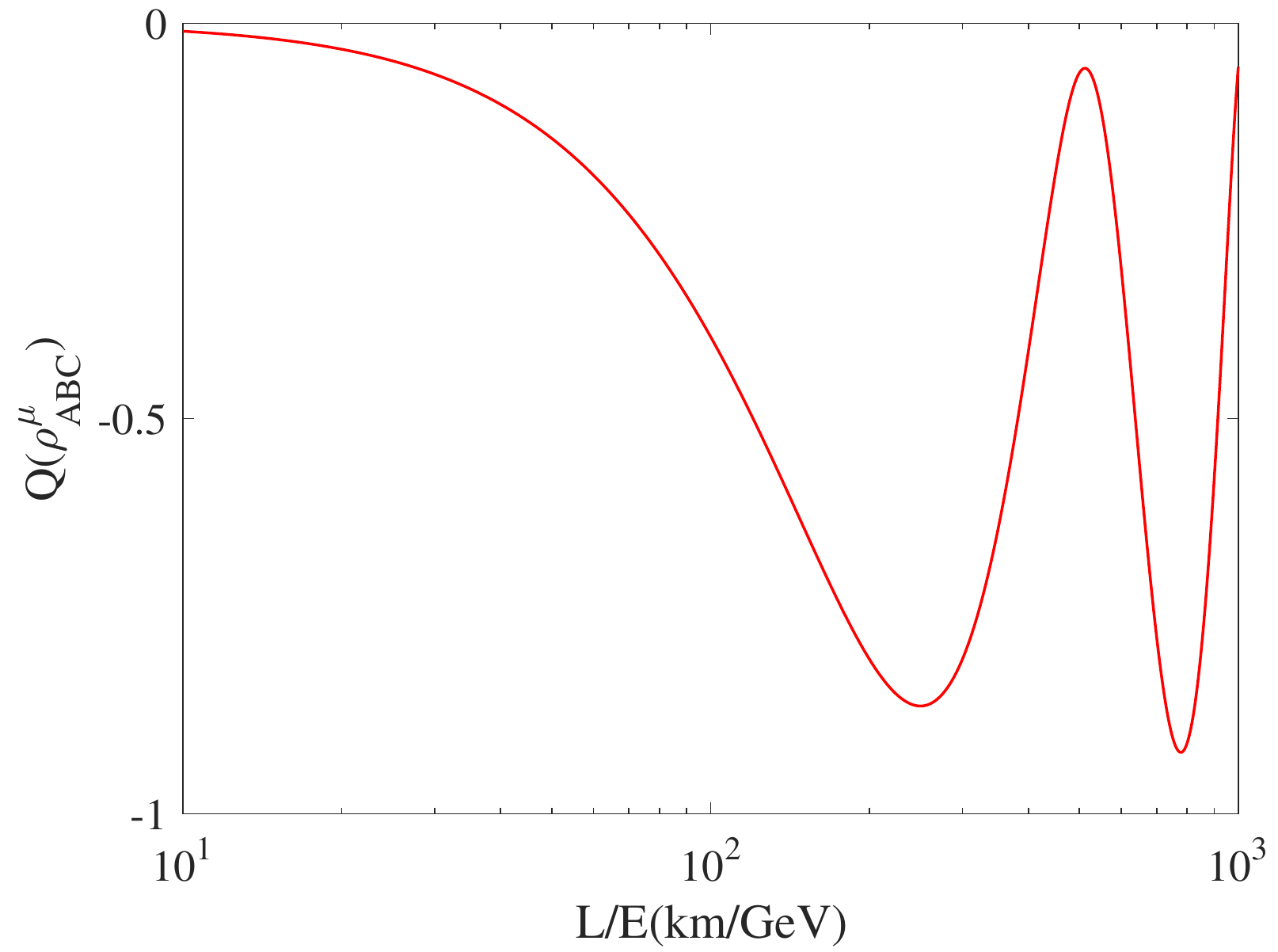}
\caption{The additivity relation $\left[C\left( {{\rho^\mu _{AC}}} \right) + C\left( {{\rho^\mu _{AB}}} \right) - C({\rho^\mu _{ABC}})\right]$ (red, solid line) of quantum coherence in the muon antineutrino oscillation system.  }
\label{f7}
\end{figure}

In Fig. \ref{f7} , we plot the change of $Q_{\rho^\mu _{ABC}}$ with respect to $L/E$ for the muon antineutrino oscillation system. One can see that $Q_{\rho^\mu _{ABC}}$ is less than zero in the
range $[10^1, 10^3]$ of $L/E$ with dimension $\rm km/GeV$. It indicates that the sum of the bipartite coherence is equal to or less than than the tripartite coherence in the muon antineutrino oscillation system.

\section{Conclusions}

In this paper, we have studied several trade-off relations in QRT, based on Bell-CHSH violations, first-order coherence and intrinsic concurrence, and the relative entropy of coherence,
for initial electron-neutrino and muon-neutrino oscillations.
To be specific, while the maximum violation of $\rm {CHSH}$ tests on bipartite states ${\left\langle \rm{CHSH} \right\rangle ^2}_{{\rho _{AB}}}$, ${\left\langle \rm{CHSH} \right\rangle ^2}_{{\rho _{BC}}}$, and ${\left\langle \rm{CHSH} \right\rangle ^2}_{{\rho _{AC}}}$ change differently with the ratio $L/E$ in a long range, their sum is always equivalent to or less than 12 for the electron and the muon neutrinos.
For the three-flavor neutrino oscillations, the trade-off relation between first-order coherence and intrinsic concurrence satisfy the relation that ${D^2}({\rho _{ABC}}) + 2C_I^2({\rho _{ABC}})/3 = 1$,
from which we can know the the amount of two quantum resources can be mutual converted in the neutrino propagation. Furthermore, it is shown that the sum of the
bipartite coherence is equal to or less than the tripartite
coherence in the neutrino flavor changing process. This implies that the tripartite coherence of three-flavor NOs
contains coherence of the reduced pairwise neutrino subsystems. These trade-off relations give a strong limitations on the
distribution of quantum resources, such as, nonlocality, concurrence, and quantum coherence, among the subsystems in NOs.
The results provide the good applications of QRT on quantum resource conversion and allocation in NOs in the future.

\begin{acknowledgements}
This work was supported by the National Science Foundation of China (Grant nos. 12004006, 12075001), Anhui Provincial Key Research and Development Plan (Grant No. 2022b13020004), and Anhui Provincial Natural Science Foundation (Grant no. 2008085QA43).
\end{acknowledgements}

\end{document}